\documentclass[12pt]{article}
\pdfoutput=1
\usepackage{subfigure}
\usepackage{amssymb,amsmath}
\usepackage{graphicx}
\usepackage{color}
\usepackage{cancel}
\usepackage[colorlinks=true
,urlcolor=blue
,citecolor=blue
,linkcolor=blue
,pagecolor=blue
,linktocpage=true
,pdfproducer=medialab
]{hyperref}
\usepackage[a4paper,width=15.2cm]{geometry}

\usepackage{ulem}
\usepackage[section]{placeins}

\makeatletter \renewcommand{\@dotsep}{10000} \makeatother
\def\be{\begin{equation}}
\def\ee{\end{equation}}
\def\bea{\begin{eqnarray}}
\def\eea{\end{eqnarray}}
\def\bi{\begin{itemize}}
\def\ei{\end{itemize}}

\usepackage[nodisplayskipstretch]{setspace}

\newcommand{\beq}{\begin{equation}}
\newcommand{\eeq}{\end{equation}}

\newcommand{\eval}{\biggr\rvert}
\begin{document}

\date{\today}

\begin{center}
{\Large\bf 
Quasi Yukawa Unification and Fine-Tuning in $\mathbf{U(1)}$ Extended SSM } 
\end{center}

\begin{center}

{\Large
Ya\c{s}ar Hi\c{c}y\i lmaz$^{a}$, Meltem Ceylan$^{a}$, Asl\i~ Alta\c{s}$^{a}$

\vspace{0.3cm}
Levent Solmaz$^{a}$ and Cem Salih \"{U}n$^{b,c}$
}

\vspace{0.75cm}

{\it \hspace{-2.4cm}$^a$
Department of Physics, Bal\i kesir University, 10145, Bal\i kesir Turkey} \\

\vspace{0.2cm}
{\it $^b$
Center of Fundamental Physics, Zewail City of Science and Technology, 6 October City, 12588, Cairo, Egypt} \\

\vspace{0.2cm}
 {\it \hspace{-3.1cm}$^c$
Department of Physics, Uluda{\~g} University, 16059, Bursa, Turkey
}

\vspace{1.5cm}
\section*{Abstract}
\end{center}
\noindent

We consider the low scale implications in the $U(1)'$ extended MSSM (UMSSM). We restrict the parameter space such that the lightest supersymmetric particle (LSP) is always the lightest neutralino. In addition, we impose quasi Yukawa unification (QYU) at the grand unification scale ($M_{{\rm GUT}}$). QYU strictly requires the ratios among the yukawa couplings as $y_{t}/y_{b}\sim 1.2$, $y_{\tau}/y_{b}\sim 1.4$, and $y_{t}/y_{\tau}\sim 0.8$. We find that the need of fine-tuning over the fundamental parameter space of QYU is in the acceptable range ($\Delta_{EW} \leq 10^{3}$), even if the universal boundary conditions are imposed at $M_{{\rm GUT}}$, in contrast to CMSSM and NUHM. UMSSM with the universal boundary conditions yields heavy stops ($m_{\tilde{t}} \gtrsim 2.5$ TeV), gluinos ($m_{\tilde{g}} \gtrsim 2$ TeV), and squarks from the first two families ($m_{\tilde{q}} \gtrsim 4$ TeV). Similarly the stau mass is bounded from below at about $1.5$ TeV. Despite this heavy spectrum, we find $\Delta_{EW} \gtrsim 300$, which is much lower than that needed for the minimal supersymmetric models. In addition, UMSSM yield relatively small $\mu-$term, and the LSP neutralio is mostly form by the Higgsinos of mass $\gtrsim 700$ GeV. We obtain also bino-like dark matter (DM) of mass about 400 GeV. Wino is usually found to be heavier than Higgsinos and binos, but there is a small region where $\mu \sim M_{1}\sim M_{2}\sim 1$ TeV. We also identify chargino-neutralino coannihilation channel and $A-$resonance solutions which reduce the relic abundance of LSP neutralino down to the ranges compatible with the current WMAP measurements.

\newpage

\renewcommand{\thefootnote}{\arabic{footnote}}
\setcounter{footnote}{0}



\section{Introduction}
\label{sec:intro}

Even if the minimal supersymmetric extension of the Standard Model (MSSM) is compatible with the current experimental measurements for the Higgs boson, recent studies show that realizing  a Higgs boson of mass around 125 GeV in minimal  models such as constrained MSSM (CMSSM) and models with non-universal Higgs masses (NUHM) requires  a heavy supersymmetric particle spectrum. The Higgs boson of mass about 125 GeV leads to the stop quark mass in multi-TeV range \cite{Ajaib:2012vc}, or necessitates a large soft supersymmetry breaking (SSB) trilinear term $A_{t}$ \cite{Carena:2011aa}. In addition to the Higgs boson results, also absence of a direct signal in the experiments conducted in the Large Hadron Collider (LHC) has lifted up the mass bounds on the supersymmetric particles, especially in the color sector. For instance, the current results exclude the gluino of mass lighter than $\sim 1.8$ TeV when $m_{\tilde{g}} \ll m_{\tilde{q}}$ \cite{gluinoATLAS}, which becomes severer when $m_{\tilde{g}}\simeq m_{\tilde{q}}$, where $\tilde{q}$ denotes the squarks from the first two families. Even though these bounds are mostly for R-parity conserved CMSSM, they are applicable for a large class of supersymmetric models. 

While there are numerous motivations behind the supersymmetry (SUSY) searches, such heavy spectrum has brought naturalness under scrutiny. It is clear that the recent experimental constraints cannot be satisfied in the natural region identified with $m_{\tilde{t}_{1}},m_{\tilde{t}_{2}},m_{\tilde{b}_{1}} \lesssim 500$ GeV \cite{Kitano:2005wc}. Even though it is possible to find $m_{\tilde{t}_{1}} \ll 500$ GeV \cite{Demir:2014jqa}, $m_{\tilde{t}_{2}}$ needs to be very heavy because of the necessity of large mixing. Apart from the natural region, one might measure how much fine-tuning is required by considering the $Z-$boson mass ($M_{Z} =91.2 $ GeV)

\begin{equation}
\frac{1}{2}M_{Z}^{2} = -\mu^{2}+\frac{(m_{H_{d}}^{2}+\Sigma_{d}^{d})-(m_{H_{u}}^{2}+\Sigma_{u}^{u})\tan^{2}\beta}{\tan^{2}\beta -1}
\label{zmass}
\end{equation}
where $\mu$ is the bilinear mixing of the MSSM Higgs doublets, $\tan\beta \equiv \langle H_{u} \rangle/\langle H_{d} \rangle$, ratio of vacuum expectation values (VEVs), $\Sigma_{u,d}^{u,d}$ are the radiative effects from the Higgs potential and $m_{H_{u,d}}^{2}$ are the soft supersymmetry breaking (SSB) mass terms for the Higgs doublets $H_{u,d}$. A recent work \cite{Baer:2012mv} has defined the following parameter to quantify the fine-tuning measure 

\begin{equation}
\Delta_{EW}\equiv {\rm Max}(C_{i})/(M_{Z}^{2}/2)
\label{DeltaEW}
\end{equation}
where
\begin{equation}
C_{i}\equiv \left\lbrace 
\begin{array}{c}
\hspace{-1.2cm}C_{H_{d}}=\mid m^{2}_{H_{d}}/(\tan^{2}\beta -1) \mid \\ \\
C_{H_{u}}=\mid m^{2}_{H_{u}}\tan^{2}\beta/(\tan^{2}\beta -1) \mid \\ \\
\hspace{-3.8cm}C_{\mu}=\mid -\mu^{2}\mid .
\end{array}
\right. 
\end{equation}

The fine-tuning can be interpreted as the presence of some missing mechanisms and its amount measures the effects of such missing mechanisms. Their effects can be reflected within SUSY models by considering non-universality or adding extra sectors to the theory \cite{Hall:2011aa}. In this respect, it is interesting to probe the models beyond the MSSM models in light of the current experimental results.

Note that in contrast to the natural region characterized with the stop and sbottom masses, the fine-tuning does not depend on these masses directly. From moderate to large $\tan\beta$ values, $\mu^{2} \approx -m_{H_{u}}^{2}$ is needed in order to obtain correct Z boson mass $M_{Z}$; hence, the fine-tuning is mostly determined by $C_{\mu}$, unless $\mu$ is so small that the large radiative corrections to $m_{H_{u}}$ are needed in Eq.(\ref{zmass}). Thus,  large stop or sbottom masses can still yield an acceptable amount of fine-tuning. The conclusion that the fundamental parameter spaces of CMSSM and NUHM need to be highly fine-tuned is raised due to the strict universality in the boundary conditions of these models. 

In this work we consider the MSSM extended by an additional $U(1)^\prime$ group (UMSSM) in the simplest form. A general extension of MSSM by a U(1) group can be realized from an underlying GUT theory involving a gauge group larger than SU$(5)$. For instance the following symmetry braking chain

\begin{equation}\hspace{-1.0cm}
E(6)\rightarrow SO(10)\times U(1)_{\psi} \rightarrow SU(5)\times U(1)_{\psi}\times U(1)_{\chi}\rightarrow G_{{\rm MSSM}}\times U(1)'
\label{e6chain}
\end{equation}
where $G_{{MSSM}}=SU(3)_{c}\times SU(2)_{L}\times U(1)_{Y}$ is the MSSM gauge group, and U(1)$^\prime$ can be expressed as a general mixing of $U(1)_{\psi}$ and $U(1)_{\chi}$ as 

\begin{equation}
U(1)'=\cos\theta_{E_{6}}U(1)_{\chi}+\sin\theta_{E_{6}}U(1)_{\psi}
\label{umix}
\end{equation}

Emergence of SO$(10)$ and/or SU$(5)$ allows one to imply a set of boundary conditions, which can be suited in these groups. For instance, the supersymmetric particle masses can be set universal at the grand unified scale ($M_{{\rm GUT}}$) in $SO(10)$, while two different mass scales can be imposed to the fields in $\mathbf{5}$ and $\mathbf{10}$ representations of $SU(5)$.

In exploring this extension, we briefly aim to analyze the effects only from having another gauge sector, which is not included in the minimal SUSY models, by imposing universal boundary conditions at $M_{{\rm GUT}}$. In addition to the boundary conditions imposed on the fundamental parameters, we also restict the Yukawa sector such that the Yukawa couplings, especially for the third family matter fields, are determined by the minimal $E(6)$ (or $SO(10)$) unification scheme. If a model based on E$(6)$ gauge group \cite{Lazarides:1986cq} is constructed in a minimal fashion in a way that all the matter fields of a family are resided in a $\mathbf{27}$ dimensional representation and the Higgs fields in another $\mathbf{27}$, such a model also proposes unification of the Yukawa couplings (YU) as well as the gauge couplings. This elegant scheme of unification can be maintained if E$(6)$ is broken down to the MSSM gauge group via SO$(10)$, since models based on the SO$(10)$ gauge group reserves YU. Even though it is imposed at $M_{{\rm GUT}}$, YU is also strongly effective at the low scale, since it requires the threshold corrections at the low scale \cite{Gogoladze:2010fu}. Relaxing YU to $b-\tau$ YU does not weaken its strength on the low scale implications, since $y_{b}$ still requires  large and negative SUSY threshold corrections \cite{Raza:2014upa}.

Despite its testable predictions at LHC \cite{Gogoladze:2010fu,bigger-422}, YU rather leads to contradictory mass relations such that $N=U \propto D=L$ and $m_{c}^{0}/m_{t}^{0} = m_{s}^{0}/m_{b}^{0}$, $m_{s}^{0}=m_{\mu}^{0}$, and $m_{d}^{0}=m_{e}^{0}$. One way to avoid this contradiction and obtain realistic fermion masses and mixing is to propose vector-like matter multiplets at the GUT scale \cite{Witten:1979nr}, which are allowed to mix with fermions in 16-plet representation of $SO(10)$. This approach is also equivalent to introduce non-renormalizable couplings along with non-zero vacuum expectation value (VEV) of a non-singlet SO$(10)$ field \cite{Anderson:1993fe}. Another way is to extend the Higgs sector with an assumption that the MSSM Higgs doublets are superpositions of fields from different SO$(10)$ representations \cite{Babu:1992ia}. 

Even though YU for the third family can be consistently maintained under assumptions that the extra fields negligibly interact with the third family and the MSSM Higgs doublets solely reside in 10 dimensional representation of SO$(10)$, these two approaches, in general, break YU in SO$(10)$. On the other hand, if one can formulate the asymptotic relation among the Yukawa couplings, then the contributions can be restricted such that the quasi-YU (QYU) can be maintained. For instance, It was shown in Ref. \cite{Gomez:2002tj} that in the presence of Higgs fields from $H'(15,1,3)$ in addition to those from $h(1,2,2)$ of the Pati-Salam model \cite{Pati:1974yy} Yukawa couplings at $M_{{\rm GUT}}$ can be expressed as 

\begin{equation}
 y_{t}:y_{b}:y_{\tau}=\mid 1+ C \mid : \mid 1- C \mid : \mid 1+ 3C \mid
 \label{QYU}
\end{equation}  

The gauge group of the Pati-Salam Model, $G_{{\rm PS}}=SU(4)_{c}\times SU(2)_{L}\times SU(2)_{R}$ is the maximal subgroup of $SO(10)$, and hence these extra Higgs fields can be employed in $SO(10)$ GUT models. The parameter $C$ denotes the contributions to Yukawa couplings from the extra Higgs fields, and restricting these contributions as $C \leq 0.2$, Eq.(\ref{QYU}) refers to the QYU condition. Note that $C$ can be either positive or negative, but it is possible to restrict it to positive values without lose of generality by adjusting the phase of the representations $H'$ and $h$. QYU yield significantly different low scale phenomenology \cite{Karagiannakis:2015rma} than the exact YU. In addition, QYU can provide an interesting scenario in respect of the fine-tuning, since a better fine-tuning prefers that the ratios of Yukawa couplings are different from unity \cite{Antusch:2011xz}, when the universal boundary conditions are imposed at $M_{{\rm GUT}}$.

In this work we analyze the fine-tuning requirements in UMSSM with QYU condition imposed at the GUT scale. The outline of the paper is the following. We will briefly describe UMSSM in Section \ref{sec:model}. After summarizing our scanning procedure and the experimental constraints we employ in our analysis in Section \ref{sec:scan}, we present our results in the fundamental parameter space of QYU in Section \ref{sec:QYUFun}. The mass spectrum of the supersymmetric particles and dark matter (DM) implications are considered in Section \ref{sec:spec}. Finally, we summarize and conclude our results in Section \ref{sec:conc}.
\section{Model Description}
\label{sec:model}

In this section, we briefly summarize the $E(6)$ based supersymmetric $U(1)^\prime$ models whose symmetry breaking patterns and resultant gauge group is given in Eq.(\ref{e6chain}) (For a detailed consideration see \cite{Barr:1985qs,Langacker:1998tc}). The superpotential in such models can be given as 

\begin{equation}
W = Y_{u}\hat{Q}\hat{H}_{u}\hat{U}^{c}+Y_{d}\hat{Q}\hat{H}_{d}\hat{D}^{c}+Y_{e}\hat{L}\hat{H}_{d}\hat{E}^{c}+h_{s}\hat{S}\hat{H}_{d}\hat{H}_{u}.
\label{suppot1}
\end{equation}
where $\hat{Q}$ and $\hat{L}$ denote the left-handed chiral superfields for the quarks and leptons, while $\hat{U}^{c}$, $\hat{D}^{c}$ and $\hat{E}^{c}$ stand for the right-handed chiral superfields of u-type quarks, d-type quarks and leptons, respectively. $H_{u}$ and $H_{d}$ MSSM Higgs doublets and $Y_{u,d,e}$ are their Yukawa couplings to the matter fields. Finally $\hat{S}$ denote a chiral superfield, which does not exist in MSSM. This field is singlet under the MSSM group and its VEV is responsible for the braking of $U(1)^\prime$ symmetry. The invariance under $U(1)^\prime$ requires an appropriate charge assignment for the MSSM fields. Table \ref{charges} displays the charge configurations for $U(1)_{\psi}$ and $U(1)_{\chi}$ models. Note that Eq. (\ref{umix}) allows infinite number of different charge configurations depending on $\theta_{E_{6}}$.

\begin{table}[ht!]
\setstretch{1.5}
\centering
\begin{tabular}{|c|c|c|c|c|c|c|c|c|}
\hline
 Model & $\hat{Q}$ & $\hat{U}^{c}$ & $\hat{D}^{c}$ & $\hat{L}$ & $\hat{E}^{c}$ & $\hat{H}_{d}$ & $\hat{H}_{u}$ & $\hat{S}$ \\
 \hline
$ 2\sqrt{6}~U(1)_{\psi}$ & 1 & 1 & 1 &1 &1 & -2 & -2 & 4 \\
\hline 
$ 2\sqrt{10}~U(1)_{\chi}$ & -1 & -1 & 3 & 3 & -1 & -2 & 2 & 0 \\
\hline  
\end{tabular}
\caption{Charge assignments for the fields in several models.}
\label{charges}
\end{table}

Eq.(\ref{suppot1}) is almost the same as the superpotential in MSSM except the last term. As is well known, a bilinear mixing of the MSSM Higgs doublets are introduced with the term $\mu \hat{H_{u}}\hat{H_{d}}$ in MSSM, and $\mu$-term plays an essential role in the electroweak symmetry breaking (EWSB). However, in MSSM, $\mu -$term preserves the SUSY, and hence; it can be at any scale, despite its connection with the EWSB. This is so-called $\mu -$problem in MSSM. On the other hand, if an extra $U(1)^\prime$ group, under which the MSSM fields have non-trivial charges, is introduced, the invariance principle forbids to introduce such terms like $\mu \hat{H_{u}}\hat{H_{d}}$, since $H_{u}$ and $H_{d}$ are charged under $U(1)^\prime$, and their charges do not have to cancel each other. Rather, another term can be introduced such as $h_{s}\hat{S}\hat{H}_{d}\hat{H}_{u}$, where $S$ is a dynamical field, and its non-zero VEV breaks the extra $U(1)^\prime$ symmetry, while it also induces a bilinear mixing between $H_{u}$ and $H_{d}$ with $\mu \equiv h_{s}\langle S \rangle$. In this picture, the $\mu-$term can be related to $U(1)^\prime$ breaking scale and it can be generated dynamically. 

Before proceeding, one of the important task about $U(1)^\prime$ models is to deal with the anomalies and make sure that the model under concern is anomaly free. There are several attempts \cite{Cheng:1998nb} by either adding exotics or imposing non-universal charges to the families. The charge assignments given in Table \ref{charges} corresponds to the universal charge configurations for the families. In this case, one should carefully consider the exotics, since their existence may bring back the $\mu-problem$ or break the gauge coupling unification. The gauge coupling unification can be maintained if another ($\mathbf{27}_{L}+\mathbf{27}_{L}^{*}$) is added and assumed to yield only MSSM Higgs-like doublets can be light \cite{Langacker:1998tc}. Even if the exotics are heavy and they decouple at a high energy in compared to $M_{{\rm GUT}}$, they can still contribute to the proton decay. In this case, one can consider UMSSM along with $SO(10)$ which forbids baryon and lepton number violating processes \cite{Cheng:1998nb}. Finally we should note that the existence of right-handed neutrinos. We neglect the contributions from the right-handed neutrinos, since these contributions are suppressed due to smallness of the established neutrino masses \cite{Wendell:2010md}, unless the inverse seesaw mechanism is imposed \cite{Mohapatra:1986bd}. 

In addition to the MSSM particle content, UMSSM yields two more particles at the low scale, one of which is the gaugino associated with the gauge fields of $U(1)^\prime$, and the other is the supersymmetric partner of the MSSM singlet $S$. Since these two particles are of no electric charge, they mix with the MSSM neutralinos after EWSB, which enriches the dark matter implications in UMSSM \cite{Frank:2014bma}. EWSB also yields a mixing $Z-Z'$, where $Z'$ is the gauge boson associated with $U(1)^\prime$. Hence, one can expect some effects from interference of $Z'$, but since the mass bound on $Z'$ is strict, these effects are highly suppressed by its heavy mass. Finally, the content of the charged sector of MSSM remain the same, but $h_{s}$ and $\langle S \rangle$ are effective in this sector, since they generate the $\mu-$term effectively, which also determined the mass of higgsinos.

A minimal $E(6)$ model, in which the matter fields are resided in $\mathbf{27}-$plet, and the MSSM Higgs fields in ($\mathbf{27}_{L}+\mathbf{27}_{L}^{*}$), also proposes YU via $y~ \mathbf{27}_{i}\mathbf{27}_{j}\mathbf{27}_{H}$ in the superpotential. The discussion on the contradictory mass relations in the fermion sector can be handled by extending the Higgs sector with $\mathbf{351}$ and $\overline{\mathbf{351}}$-plets within the $E(6)$ framework \cite{Babu:2015psa}. In our work, we assume the minimal lay out for the $E(6)$ model. However, the Higgs fields emerging from ($\mathbf{27}_{L}+\mathbf{27}_{L}^{*}$) can also break $SO(10)$ to the Pati-Salam model \cite{Howl:2007zi} which is based on the gauge group $G_{{\rm PS}} \equiv SU(4)_{c}\times SU(2)_{L}\times SU(2)_{R}$. In such a framework, the Yukawa sector may include also interactions between the matter fields and the Higgs fields from $H'(15,1,3)$ representation of $G_{{\rm PS}}$. If one assumes that $G_{{\rm PS}}$ breaks into the MSSM gauge group at about the GUT scale, the known Yukawa couplings can be stated as given in Eq.(\ref{QYU}) at $M_{{\rm GUT}}$. Note that emergence of $G_{{\rm PS}}$ in the breaking chain allows non-universal gaugino masses at the GUT scale such that \cite{Gogoladze:2009ug}

\begin{equation}
M_{1}=\dfrac{3}{5}M_{2}+\dfrac{2}{5}M_{3}
\label{PSgauginos}
\end{equation}

In this case, $M_{3}$ can be varied over the parameter space as a free parameter, and hence, the tension from the heavy gluino mass bound can be significantly relaxed, which yield drastic improvement in regard of the fine-tuning. However, as stated above, we restrict ourselves with the universal boundary conditions, and we will impose only one SSB mass term for all the three gauginos.

A solution can be analyzed if it is consistent with QYU or not by considering a parameter defined as 

\begin{equation}
R = \frac{{\rm Max}(C_{1},C_{2},C_{3})}{{\rm Min}(C_{1},C_{2},C_{3})}
\label{rquasi}
\end{equation}
where 

\begin{equation}
C_{1} = \eval \frac{y_{t}-y_{b}}{y_{t}+y_{b}} \eval~,\hspace{0.3cm} C_{2}=\eval \frac{y_{\tau}-y_{t}}{3y_{t}-y_{\tau}}\eval~,\hspace{0.3cm} C_{3} =\eval \frac{y_{\tau}-y_{b}}{3y_{b}+y_{\tau}}\eval
\end{equation}
where $y_{t,b,\tau}$ are Yukawa couplings at $M_{{\rm GUT}}$, and $C_{1,2,3}$ denote the contributions to these couplings. The consistency with QYU requires $C_{1}=C_{2}=C_{3}$, i.e $R=1$. However, Yukawa couplings can receive some contributions from the interference of $S$, $Z'$ \cite{Un:2016hji} and even exotics at $M_{{\rm GUT}}$ as well as unknown threshold corrections from the symmetry breaking. Even though these contributions can be neglected, we allow utmost $10\%$ uncertainty in $R$ to count for such contributions. Hence, a solution compatible with QYU satisfies $R\leq 1.1$ as well as $\mid C \mid \leq 0.2$.

\section{Scanning Procedure and Experimental Constraints}
\label{sec:scan}

We have employed SPheno 3.3.3 package \cite{Porod:2003um} obtained with SARAH 4.5.8 \cite{Staub:2008uz}. In this package, the weak scale values of the gauge and Yukawa couplings presence in UMSSM are evolved to the unification scale $M_{{\rm GUT}}$ via the renormalization group equations (RGEs). $M_{{\rm GUT}}$ is determined by the requirement of the gauge coupling unification through their RGE evolutions. Note that we do not strictly enforce the unification condition $g_{1}=g_{2}=g_{3}$ at $M_{{\rm GUT}}$ since a few percent deviation from the unification can be assigned to unknown GUT-scale threshold corrections \cite{Hisano:1992jj}. With the boundary conditions given at $M_{{\rm GUT}}$, all the SSB parameters along with the gauge and Yukawa couplings are evolved back to the weak scale. Note that the gauge coupling associated with the $B-L$ symmetry is determined by the unification condition at the GUT scale by imposing $g_{1}=g_{2}=g'\approx g_{3}$, where $g'$ is the gauge coupling associated with the $U'(1)$ gauge group.

We have performed random scans over the following parameter space

\begin{equation}
\setstretch{1.2}
\begin{array}{ccc}
0 \leq & m_{0} & \leq 5 ~{\rm (TeV)} \\
0 \leq & M_{1/2} & \leq 5 ~{\rm (TeV)} \\
35 \leq & \tan\beta & \leq 60 \\
-3 \leq & A_{0}/m_{0} & \leq 3 \\
-1 \leq & A_{h_{s}} & \leq 15 ~{\rm (TeV)} \\
1 \leq & v_{s} & \leq 25~{\rm (TeV)} 
\end{array}
\label{paramSP}
\end{equation}  
where $m_{0}$ is the universal SSB mass term for all the scalar fields including $H_{u}$, $H_{d}$, $S$ fields, and similarly $M_{1/2}$ is the universal SSB mass term for the gaugino fields including one associated with $U(1)'$ gauge group. $\tan\beta=\langle v_{u} \rangle / \langle v_{d} \rangle$ is the ratio of VEVs of the MSSM Higgs doublets, $A_{0}$ is the SSB trilinear scalar interaction term. Similarly, $A_{h_{s}}$ is the SSB interaction between the $S$ and $H_{u,d}$ fields, which is varied free from $A_{0}$ in our scans. Finally, $v_{s}$ denotes the VEV of $S$ fields which indicates the $U(1)'$ breaking scale. Recall that the $\mu-$term of MSSM is dynamically generated such that $\mu = h_{s}v_{s}$. Its sign is assigned as a free parameter in MSSM, since REWSB condition can determine its value but not sign. On the other hand, in UMSSM, it is forced to be positive by $h_{s}$ and $v_{s}$. Finally, we set the top quark mass to its central value ($m_{t} = 173.3$ GeV) \cite{Group:2009ad}. Note that the sparticle spectrum is not too sensitive in one or two sigma variation in the top quark mass \cite{Gogoladze:2011db}, but it can shift the Higgs boson mass by $1-2$ GeV  \cite{Gogoladze:2011aa}.

The requirement of radiative electroweak symmetry breaking (REWSB) \cite{Ibanez:1982fr} puts an important theoretical constraint on the parameter space. Another important constraint comes from the relic abundance of the stable charged particles \cite{Nakamura:2010zzi}, which excludes the regions where charged SUSY particles such as stau and stop become the lightest supersymmetric particle (LSP). In our scans, we allow only the solutions for which one of the neutralinos is the LSP and REWSB condition is satisfied.

In scanning the parameter space, we use our interface, which employs Metropolis-Hasting algorithm described in \cite{Belanger:2009ti}. After collecting the data, we impose the mass bounds on all the sparticles \cite{Agashe:2014kda}, and the constraint from the rare B-decays such as $B_{s}\rightarrow \mu^{+}\mu^{-}$ \cite{Aaij:2012nna}, $B_{s}\rightarrow X_{s}\gamma$ \cite{Amhis:2012bh}, and $B_{u}\rightarrow \tau \nu_{\tau}$ \cite{Asner:2010qj}. In addition,  the WMAP bound \cite{Hinshaw:2012aka} on the relic abundance of neutralino LSP within $5\sigma$ uncertainty. These experimental constraints can be summarized as follows:

\begin{equation}
\setstretch{1.8}
\begin{array}{c}
m_h  = 123-127~{\rm GeV}
\\
m_{\tilde{g}} \geq 1.8~{\rm TeV} 
\\
M_{Z'} \geq 2.5 ~{\rm TeV} \\
0.8\times 10^{-9} \leq{\rm BR}(B_s \rightarrow \mu^+ \mu^-) 
  \leq 6.2 \times10^{-9} \;(2\sigma) 
\\ 
2.99 \times 10^{-4} \leq 
  {\rm BR}(B \rightarrow X_{s} \gamma) 
  \leq 3.87 \times 10^{-4} \; (2\sigma) 
\\
0.15 \leq \dfrac{
 {\rm BR}(B_u\rightarrow\tau \nu_{\tau})_{\rm MSSM}}
 {{\rm BR}(B_u\rightarrow \tau \nu_{\tau})_{\rm SM}}
        \leq 2.41 \; (3\sigma) \\
   0.0913 \leq \Omega_{{\rm CDM}}h^{2} \leq 0.1363~(5\sigma)     
\label{constraints}        
\end{array}
\end{equation}

We have emphasized the bounds on the Higgs boson\cite{:2012gk} and the gluino \cite{gluinoATLAS}, since they have drastically changed since the LEP era. Even though the mass bound on $Z'$ can be lowered through detailed analyses \cite{Accomando:2013sfa}, we require our solutions to yield heavy $Z'$, since it is not directly related to our considerations. One of the stringent bounds listed above comes from the rare B-meson decay into a muon pair, since the supersymmetric contribution to this process is proportional to $(\tan\beta)^{6}/m_{A}^{4}$. We have considered the high $\tan\beta$ region in the fundamental parameter space as given in Eq.(\ref{paramSP}), and $m_{A}$ needs to be large to suppress the supersymmetric contribution to ${\rm BR}(B_{s}\rightarrow \mu^{+}\mu^{-})$. Besides, the WMAP bound  is also highly effective to shape the parameter space, since the relic abundance of neutralino LSP is usually high over the fundamental parameter space. One needs to identify some coannihilation channels in order to have solutions compatible with the WMAP bound. The DM observables in our scan are calculated by micrOMEGAs \cite{Belanger:2006is} obtained by SARAH \cite{Staub:2008uz}. Finally, we impose the fine-tuning condition as $\Delta_{EW} \leq 10^{3}$.

\section{Fundamental Parameter Space of QYU}
\label{sec:QYUFun}

\begin{figure}[ht!]
\centering
\subfigure{\includegraphics[scale=0.405]{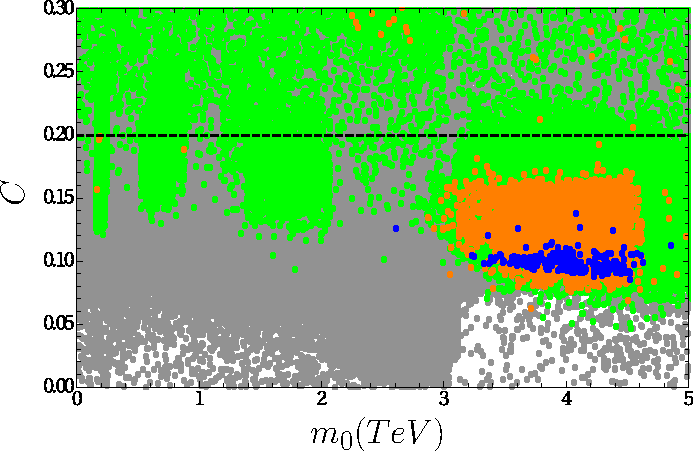}}%
\subfigure{\hspace{0.8cm}\includegraphics[scale=0.4]{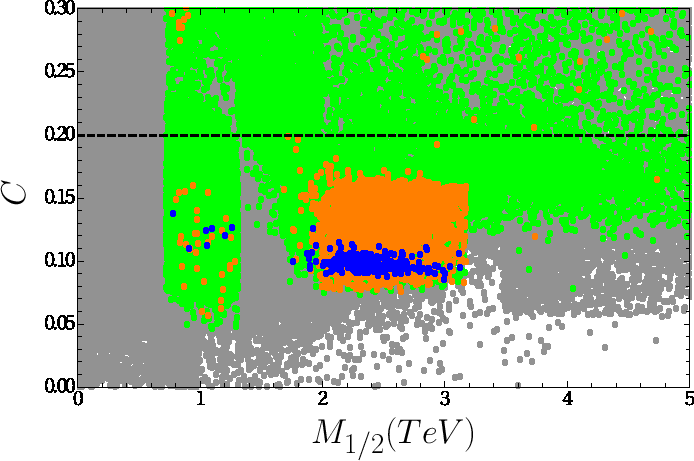}}
\subfigure{\includegraphics[scale=0.4]{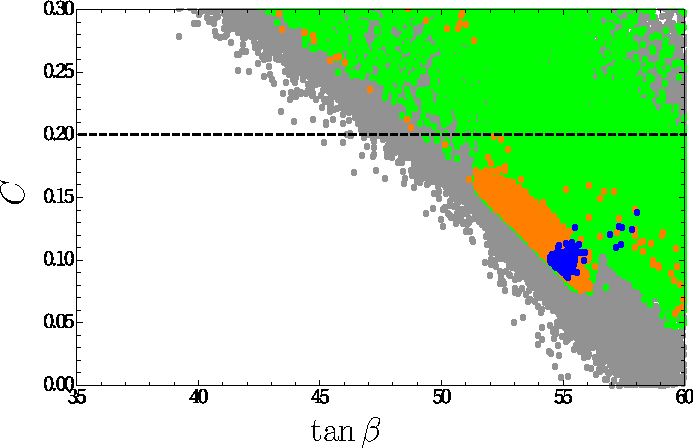}}%
\subfigure{\hspace{0.8cm}\includegraphics[scale=0.4]{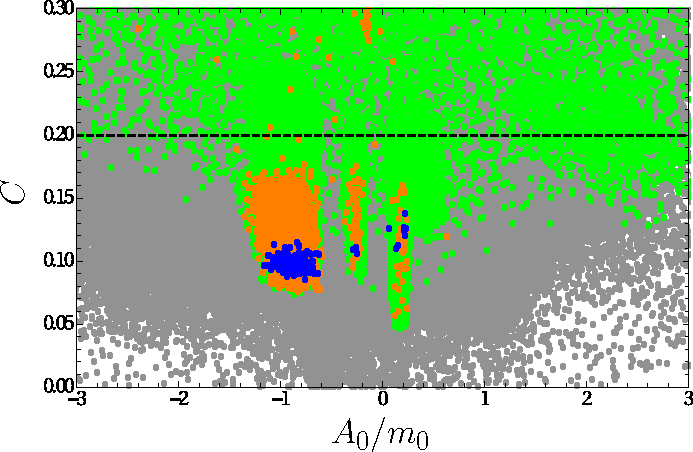}}
\caption{Plots in the $C-m_{0}$, $C-M_{1/2}$, $C-A_{0}/m_{0}$, and $C-\tan\beta$ planes. All points are consistent with REWSB and neutralino LSP. Green points satisfy the mass bounds and the constraints from the rare B-decays. Orange points form a subset of green and they are compatible with the WMAP bound on the relic abundance of neutralino LSP within $5\sigma$. Finally, the blue points are a subset of orange, which are consistent with the QYU and fine-tuning condition. The dashed lines indicates $C=0.2$.}
\label{fig1}
\end{figure}

\begin{figure}[ht!]
\centering
\subfigure{\includegraphics[scale=0.4]{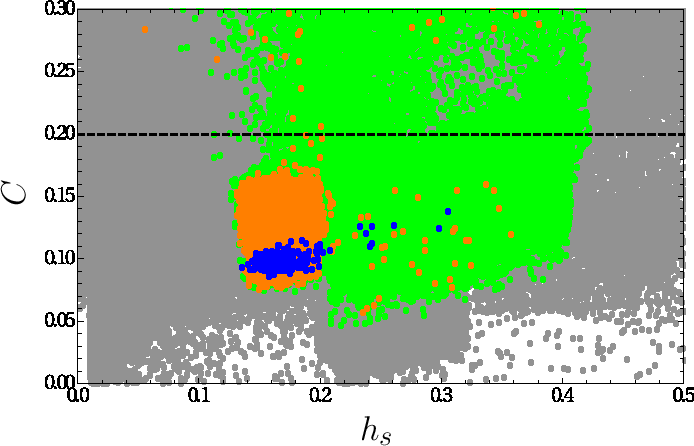}}%
\subfigure{\hspace{0.8cm}\includegraphics[scale=0.4]{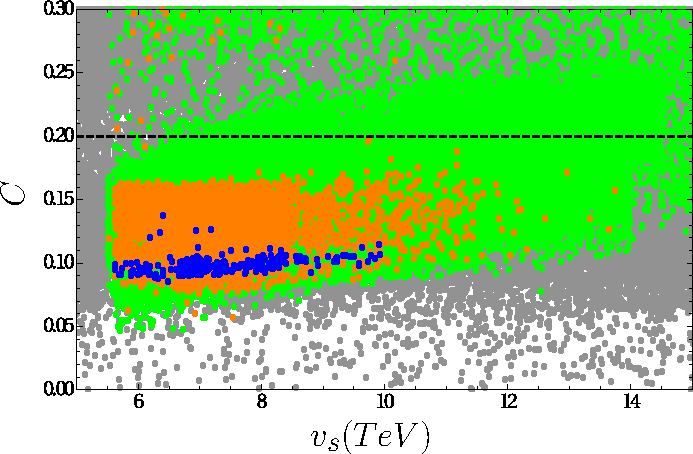}}
\caption{Plots in the $C-h_{s}$ and $C-v_{s}$. The color coding is the same as Figure \ref{fig1}.}
\label{fig2}
\end{figure}

We present our results for the fundamental parameter space in light of the experimental constraints mentioned in the previous section. Figure \ref{fig1} illustrates the QYU parameter space in a correlation with the usual CMSSM fundamental parameters in the $C-m_{0}$, $C-M_{1/2}$, $C-A_{0}/m_{0}$, and $C-\tan\beta$ planes. All points are consistent with REWSB and neutralino LSP. Green points satisfy the mass bounds and the constraints from the rare B-decays. Orange points form a subset of green and they are compatible with the WMAP bound on the relic abundance of neutralino LSP within $5\sigma$. Finally, the blue points are a subset of orange, which are consistent with the QYU and fine-tuning condition. The dashed lines indicates $C=0.2$. As seen from $C-m_{0}$, QYU requires the universal scalar mass parameter larger than 2 TeV as in the case of MSSM with non-universal gauginos imposed at $M_{{\rm GUT}}$, with the current experimental bounds $m_{0}$ is expected to be much larger in the CMSSM framework. Similarly, $C-M_{1/2}$ plane indicates that $M_{1/2}$ can only be as light as $800$ GeV. This bound is not strictly imposed by the QYU condition, rather the heavy gluino mass bound requires heavy $M_{1/2}$, when the universal gaugino masses are imposed. QYU condition mostly restrict the $\tan\beta$ parameter to the values larger than about 54 as seen from the $C-\tan\beta$ plane, as happens in the MSSM. Finally, $A_{0}$ values are mostly find in the negative region, while it is possible to realize QYU with small positive $A_{0}/m_{0}$ values.

\begin{figure}[ht!]
\centering
\subfigure{\includegraphics[scale=0.4]{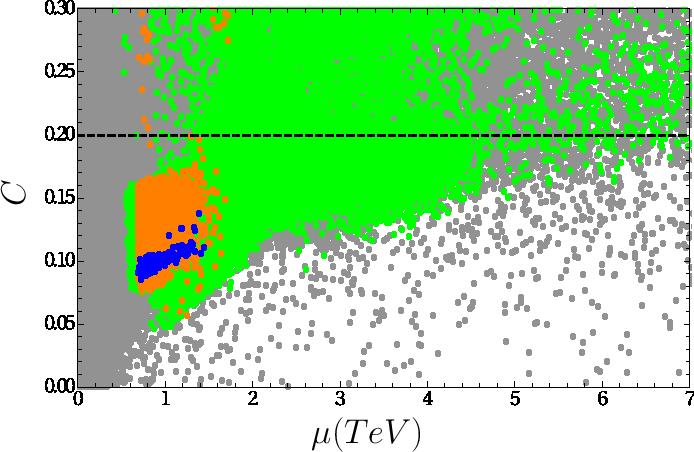}}%
\subfigure{\hspace{0.8cm}\includegraphics[scale=0.4]{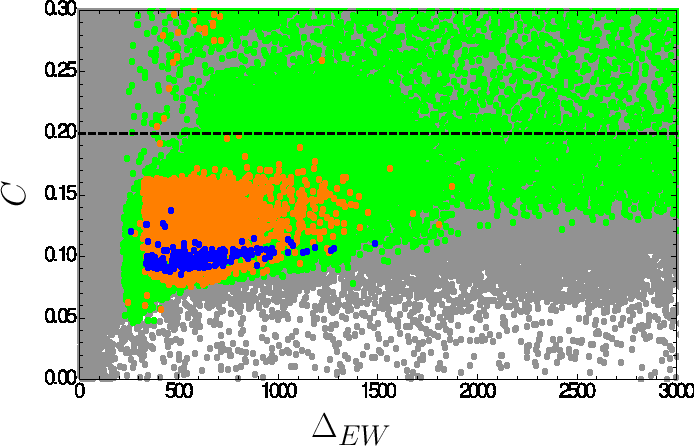}}
\subfigure{\includegraphics[scale=0.4]{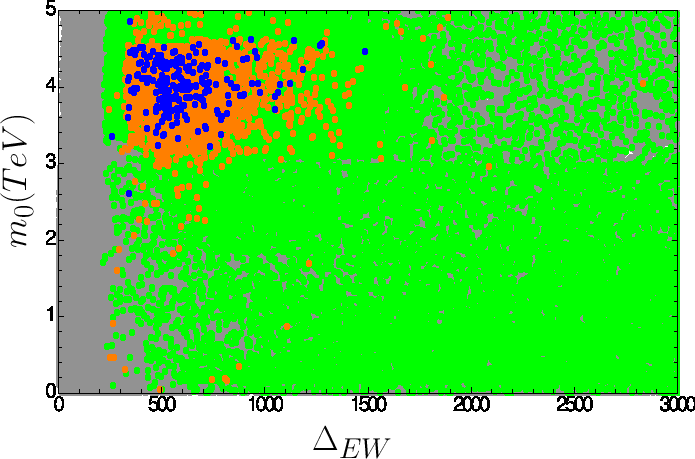}}%
\subfigure{\hspace{0.8cm}\includegraphics[scale=0.4]{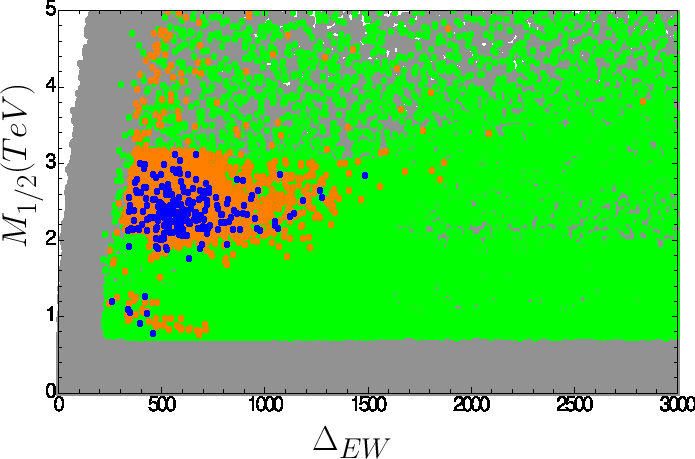}}
\caption{Plots in the $C-\mu$ and $C-\Delta_{EW}$, $m_{0}-\Delta_{EW}$, and $M_{1/2}-\Delta_{EW}$. The color coding is the same as Figure \ref{fig1} without the fine-tuning condition.}
\label{fig3}
\end{figure}

In addition to the fundamental parameters of CMSSM, Figure \ref{fig2} displays the results in UMSSM parameters with plots in the $C-h_{s}$ and $C-v_{s}$. The color coding is the same as Figure \ref{fig1}. The $C-h_{s}$ plane shows that the QYU solutions accumulate mostly in the region with $0.1 \lesssim h_{s} \lesssim 0.2$, while it can be enlarged to about 0.3 with a good statistics. On the other hand, the region with $h_{s} \gtrsim 0.4$ is excluded by the current experimental constraints (green). The plane $C-v_{s}$ shows that the lowest scale for the $U(1)'$ breaking is about 5 TeV. This breaking scale is restricted to $\lesssim 10$ TeV by QYU and the fine-tuning condition (blue).

\begin{figure}[ht!]
\centering
\subfigure{\includegraphics[scale=0.4]{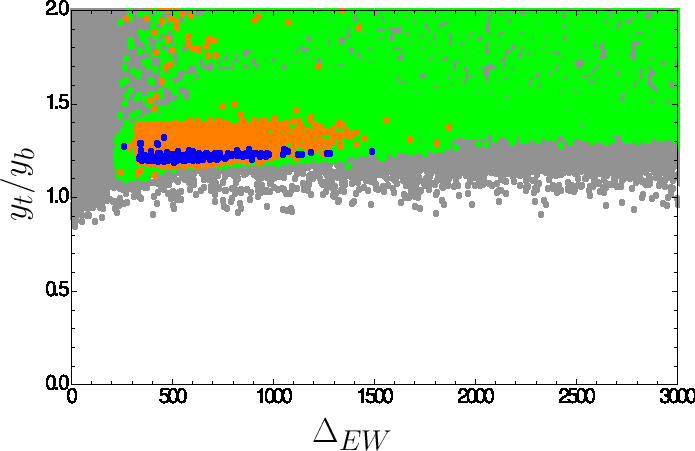}}%
\subfigure{\hspace{0.8cm}\includegraphics[scale=0.4]{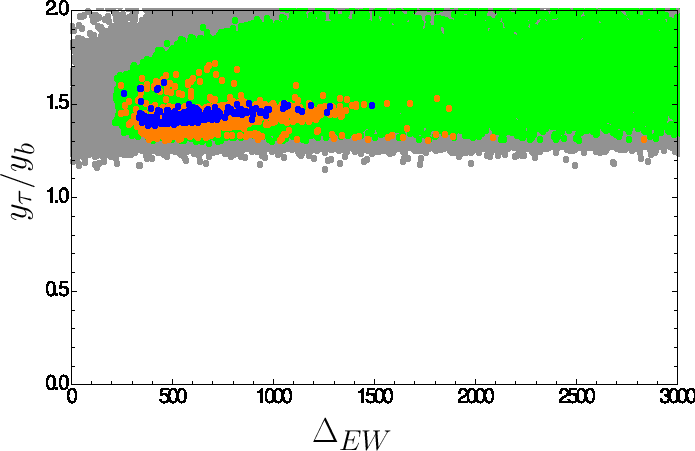}}
\subfigure{\includegraphics[scale=0.4]{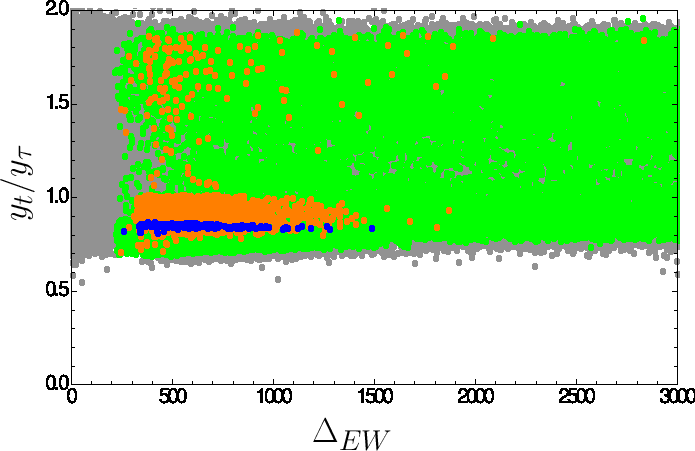}}%
\subfigure{\hspace{0.8cm}\includegraphics[scale=0.4]{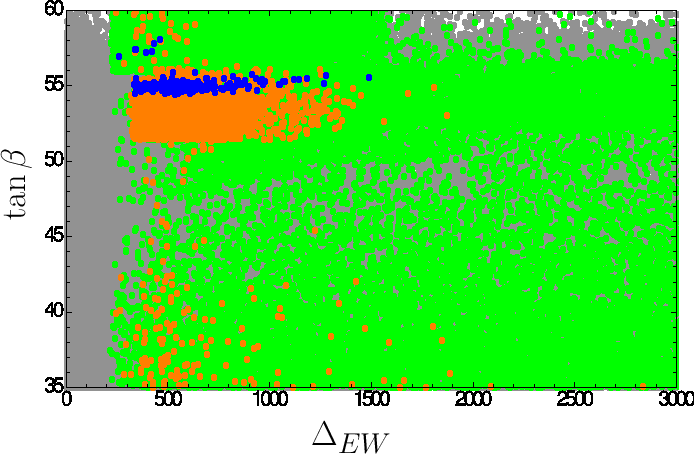}}
\caption{Plots in the $y_{t}/y_{b}-\Delta_{EW}$, $y_{\tau}/y_{b}-\Delta_{EW}$, $y_{t}/y_{\tau}-\Delta_{EW}$, and $\Delta_{EW}-\tan\beta$ planes. The color coding is the same as Figure \ref{fig1} without the fine-tuning condition.}
\label{fig4}
\end{figure}

Since, the breaking scale along with $h_{s}$ generates the $\mu -$term, it is worth to consider how large a $\mu -$term can be realized in UMSSM. Figure \ref{fig3} represents our results with plots the $C-\mu$, $C-\Delta_{EW}$, $m_{0}-\Delta_{EW}$, and $M_{1/2}-\Delta_{EW}$. The color coding is the same as Figure \ref{fig1}. The $C-\mu$ plane shows that the alignment between $v_{s}$ and $h_{s}$ allows the range $\mu \in\sim 800-1500$ GeV, which yields low fine-tuning ($\Delta_{EW}\gtrsim 300$) compatible with the QYU condition as seen from the $C-\Delta_{EW}$ plane. Such a low fine-tuning can be achieved even when $m_{0} \gtrsim 3$ TeV and $M_{1/2}\gtrsim 2$ TeV as shown in the bottom panels of Figure \ref{fig3}.

Figure \ref{fig4} displays the ratios of the Yukawa couplings with plots in $y_{t}/y_{b}-\Delta_{EW}$, $y_{\tau}/y_{b}-\Delta_{EW}$, $y_{t}/y_{\tau}-\Delta_{EW}$, and $\Delta_{EW}-\tan\beta$ planes. The color coding is the same as Figure \ref{fig1}. QYU requires certain ratios among the Yukawa couplings. Even though $y_{t}/y_{b}$ can lie from $1.1$ to about $2$, QYU rather restricts this ratio as $y_{t}/y_{b}\sim 1.2$. Similarly, it restricts $y_{\tau}/y_{b}\sim 1.4$ and $y_{t}/y_{\tau}\sim 0.8$ as seen from the $y_{\tau}/y_{b}-\Delta_{EW}$ and $y_{t}/y_{\tau}-\Delta_{EW}$ planes. These ratios hold for any value of the fine-tuning parameter. Finally the $\Delta_{EW}-\tan\beta$ indicates that $\tan\beta$ can be as high as 58 without disturbing the Yukawa coupling ratios and raising amount of the fine-tuning.

\section{Sparticle Spectrum} 
\label{sec:spec}

\begin{figure}[h!]
\centering
\subfigure{\includegraphics[scale=0.4]{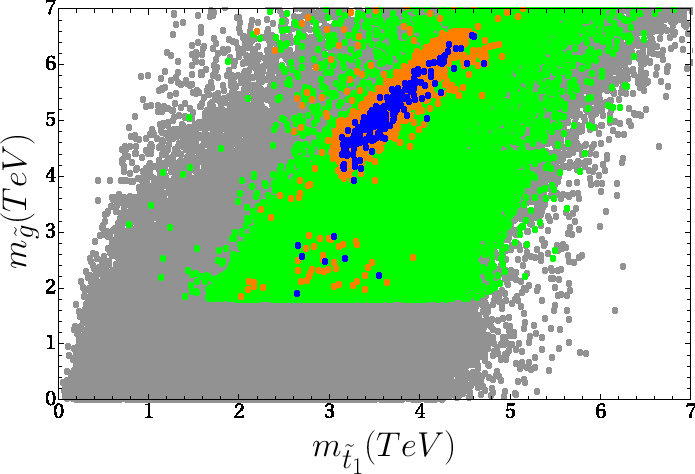}}%
\subfigure{\hspace{0.8cm}\includegraphics[scale=0.4]{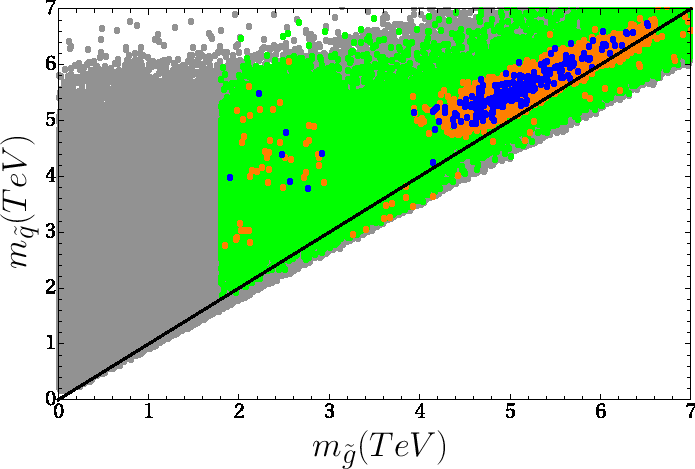}}
\subfigure{\includegraphics[scale=0.4]{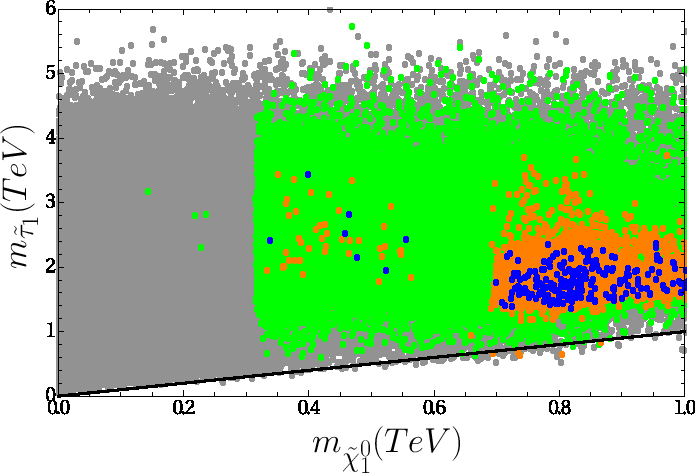}}%
\subfigure{\hspace{0.8cm}\includegraphics[scale=0.4]{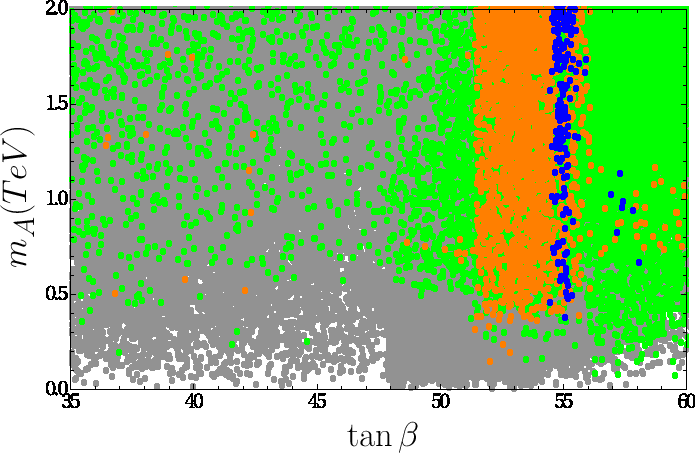}}
\caption{Plots in the $m_{\tilde{g}}-m_{\tilde{t}_{1}}$, $m_{\tilde{q}}-m_{\tilde{g}}$, $m_{\tilde{\tau}_{1}}-m_{\tilde{\chi}_{1}^{0}}$, and $m_{A}-\tan\beta$ planes. The color coding is the same as Figure \ref{fig1}.}
\label{fig5}
\end{figure}

This section will present the sparticle spectrum compatible with QYU. We start with the color sector as well as staus and $m_{A}$ as shown in Figure \ref{fig5} with plots in the $m_{\tilde{g}}-m_{\tilde{t}_{1}}$, $m_{\tilde{q}}-m_{\tilde{g}}$, $m_{\tilde{\tau}_{1}}-m_{\tilde{\chi}_{1}^{0}}$, and $m_{A}-\tan\beta$ planes. The color coding is the same as Figure \ref{fig1}. As seen from the $m_{\tilde{g}}-m_{\tilde{t}_{1}}$ plane, the gluino can be as light as about 2 TeV, while the region with $m_{\tilde{t}_{1}} \lesssim 2.5$ TeV is not compatible with the QYU condition. Even though it is possible to realize the Higgs boson of mass about 125 GeV with light stops in the UMSSM framework, such light stop solutions are mostly excluded by the heavy gluino mass spectrum. Similarly, the squarks from the first two families are required to be heavier than about 3 TeV as seen from the $m_{\tilde{q}}-m_{\tilde{g}}$ plane. The $m_{\tilde{\tau}_{1}}-m_{\tilde{\chi}_{1}^{0}}$ plane shows that even though one can realize the stau mass almost degenerate with the LSP neutralino consistently with the WMAP bound (orange), the QYU together with the fine-tuning condition requires $m_{\tilde{\tau}_{1}} \gtrsim 1.5$ TeV. The last panel of Figure \ref{fig5} shows  $m_{A}$ in a correlation with the $\tan\beta$ parameter. The results in the $m_{A}-\tan\beta$ plane shows that $A-$boson can be as light as 400 GeV compatible with QYU, despite the high $\tan\beta$ values. Even if one can also impose a mass bound as $m_{A} \gtrsim 800$ for high $\tan\beta$ values, there are still a significant number of solutions compatible with QYU and escape from this bound. 

\begin{figure}[ht!]
\centering
\subfigure{\includegraphics[scale=0.4]{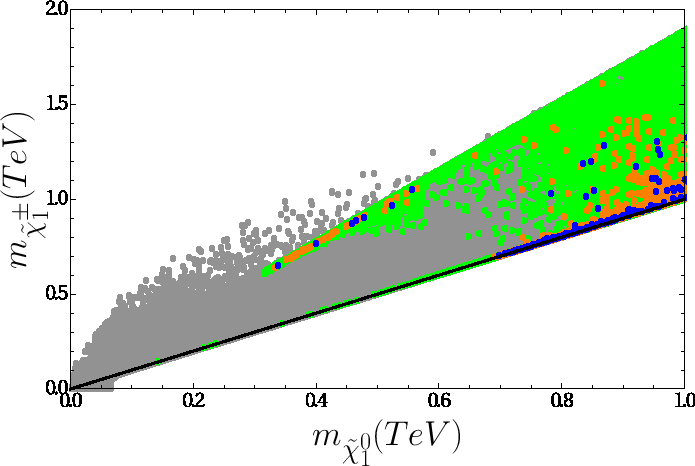}}%
\subfigure{\hspace{0.8cm}\includegraphics[scale=0.4]{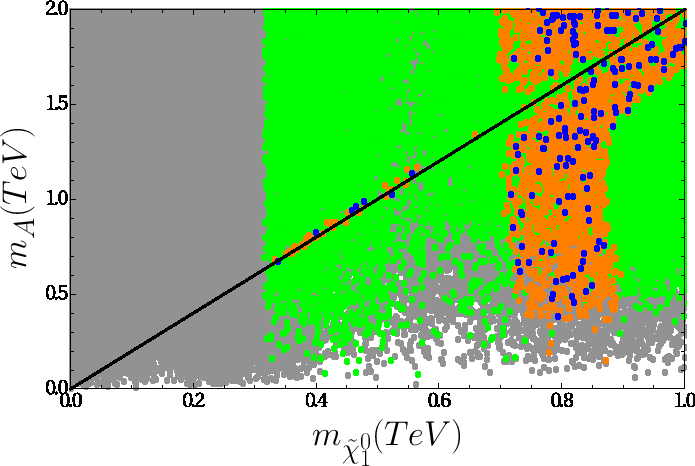}}
\caption{Plots in the $m_{\tilde{\chi}_{1}^{\pm}}-m_{\tilde{\chi}_{1}^{0}}$ and $m_{A}-m_{\tilde{\chi}_{1}^{0}}$ planes. The color coding is the same as Figure \ref{fig1}. Diagonal line indicates the region where the plotted sparticles are degenerate in mass.}
\label{fig6}
\end{figure}

Figure \ref{fig6} displays the sparticle spectrum in the $m_{\tilde{\chi}_{1}^{\pm}}-m_{\tilde{\chi}_{1}^{0}}$ and $m_{A}-m_{\tilde{\chi}_{1}^{0}}$ planes. The color coding is the same as Figure \ref{fig1}. Diagonal line indicates the region where the plotted sparticles are degenerate in mass. The $m_{\tilde{\chi}_{1}^{\pm}}-m_{\tilde{\chi}_{1}^{0}}$ shows that the chargino and LSP neutralino are mostly degenerate in mass in the region where $m_{\tilde{\chi}_{1}^{0}} \gtrsim 700$ GeV. This region may indicate the higssino DM, and the degeneracy can arise from the degeneracy of two Higgsinos. These solutions favors the chargino-neutralino coannihilation processes which reduce the relic abundance of LSP neutralino such that the solutions can be consistent with the WMAP bound. This region also yields $A-$resonance solutions as seen from the $m_{A}-m_{\tilde{\chi}_{1}^{0}}$ plane. it is also possible to realize lighter LSP neutralino solutions ($m_{\tilde{\chi}_{1}^{0}} \gtrsim 400$ GeV). There is no mass degeneracy between the LSP neutralino and chargino in this region. Hence, one can conclude the light LSP neutralino region that the LSP neutralino is Bino-like, and the WMAP bound on the relic abundance of LSP neutralino is satisfied through $A-$resonance solutions, in which two neutralinos annihilate into an $A-$boson.

\begin{figure}[ht!]
\centering
\subfigure{\includegraphics[scale=0.4]{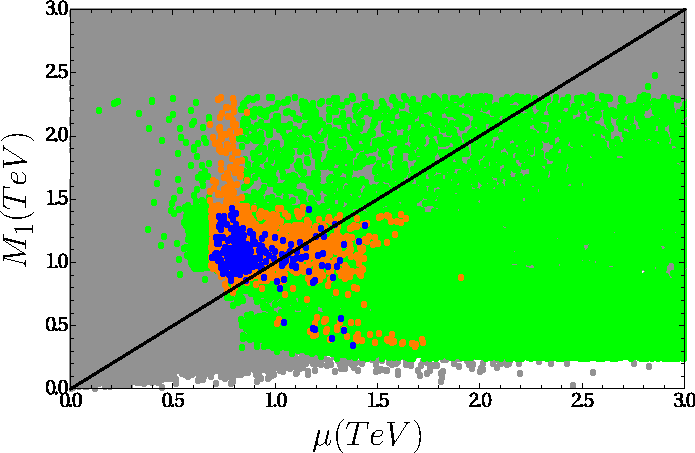}}%
\subfigure{\hspace{0.8cm}\includegraphics[scale=0.4]{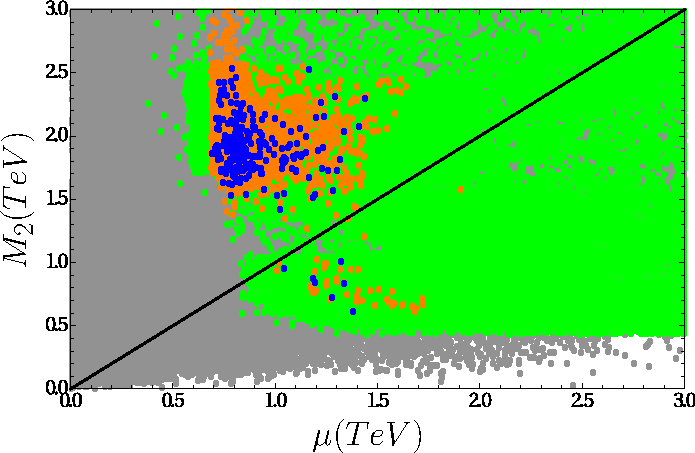}}
\caption{Plots in the $\mu-M_{1}$ and $\mu-M_{2}$ planes. The color coding is the same as Figure \ref{fig1}. The diagonal line indicates the region where $\mu=M_{1}$ ($\mu=M_{2}$) in the left (right) plane.}
\label{fig7}
\end{figure}

The LSP neutralino composition can be seen better from the $\mu-M_{1}$ and $\mu-M_{2}$ planes shown in Figure \ref{fig7}. The color coding is the same as Figure \ref{fig1}. The diagonal line indicates the region where $\mu=M_{1}$ ($\mu=M_{2}$) in the left (right) plane. The $\mu-M_{1}$ plane shows that the $\mu-$parameter is smaller than $M_{1}$ over most of the parameter space. The LSP neutralino is formed by the Higgsinos in this region. Such solutions also yield high scattering cross-section at the nuclei used in the direct detection experiments, since the interactions between quarks in nuclei and the LSP neutralino happen via Yukawa interactions. The Higgsinos and Bino are almost degenerate in mass in the region around the diagonal line, and this region indicate bino-Higgsino mixing in formation of the LSP nutralino. It is also possible to realize bino-like DM as represented with the solutions below the diagonal line where $M_{1} \leq \mu$. One can also check if it is possible to have wino mixture in formation of the LSP neutralino. The $\mu-M_{2}$ plane shows that wino is usually heavier than $\mu$ and hence $M_{1}$. However, there could be some solutions at about $\mu \sim 1$ TeV, for which the Higgsinos and wino are nearly degenerate in mass. Comparing with solutions shown in the $\mu-M_{1}$ plane, $M_{1}$ is seen to be at about 1 TeV for this solutions; i.e. $\mu \sim M_{1} \sim M_{2}$, and wino mixture in the formation of the LSP neutralino becomes as significant as the bino and higgsinos.

\section{Conclusion}
\label{sec:conc}

We explore the low scale implications in the $U(1)'$ extended MSSM (UMSSM). We restrict the parameter space such that the lightest supersymmetric particle (LSP) is always the lightest neutralino. In addition, we impose quasi Yukawa unification (QYU) at the grand unification scale ($M_{{\rm GUT}}$). The fundamental parameters of UMSSM are found to be in a large range such as $m_{0} \gtrsim 3$ TeV, $M_{1/2} \gtrsim 800$ GeV. The $\tan\beta $ parameter is mostly restricted to the region where $\tan\beta \geq 54$ by the QYU condition. Also, QYU strictly requires the ratios among the yukawa couplings as $y_{t}/y_{b}\sim 1.2$, $y_{\tau}/y_{b}\sim 1.4$, and $y_{t}/y_{\tau}\sim 0.8$. In addition, the breaking of $U(1)'$ group takes a place at the energy scales from about 5 TeV to 10 TeV.

We find that the need of fine-tuning over the fundamental parameter space of QYU is in the acceptable range, even if the universal boundary conditions are imposed at $M_{{\rm GUT}}$, in contrast to CMSSM and NUHM. Such a set up yields heavy stops ($m_{\tilde{t}} \gtrsim 2.5$ TeV), gluinos ($m_{\tilde{g}} \gtrsim 2$ TeV), and squarks from the first two families ($m_{\tilde{q}} \gtrsim 4$ TeV). Similarly the stau mass is bounded from below at about $1.5$ TeV. Despite this heavy spectrum, we find $\Delta_{EW} \gtrsim 300$, which is much lower than that needed for the minimal supersymmetric models. In addition, UMSSM yield relatively small $\mu-$term, and the LSP neutralio is mostly formed by the Higgsinos of mass $\gtrsim 700$ GeV. We obtain also bino-like dark matter (DM) of mass about 400 GeV. Wino is usually found to be heavier than Higgsinos and binos, but there is a small region where $\mu \sim M_{1}\sim M_{2}\sim 1$ TeV. We also identify chargino-neutralino coannihilation channel and $A-$resonance solutions which reduce the relic abundance of LSP neutralino down to the ranges compatible with the current WMAP measurements.


\vspace{0.2cm}
\textbf{Acknowledgement}

CSU would like to thank Qaisar Shafi for useful discussions about the GUT models based on the $E(6)$ gauge group. This work is supported in part by The Scientific and Technological Research Council of Turkey (TUBITAK) Grant no. MFAG-114F461 (CSU).

\end{document}